\pdfoutput=1
\documentclass[aps,prl,twocolumn,superscriptaddress, nobibnotes, nofootinbib]{revtex4-1}

\usepackage{color}%

\usepackage{graphicx}
\usepackage{footnote}

\usepackage[bookmarks=false]{hyperref}%
\usepackage{soul}

\begin{document}

\title{Results on the Spin-Dependent Scattering of Weakly Interacting Massive Particles on Nucleons from the Run 3 Data of the LUX Experiment}

\author{D.S.~Akerib} 
\affiliation{Case Western Reserve University, Department of Physics, 
10900 Euclid Ave, Cleveland, OH 44106, USA}
\affiliation{SLAC National Accelerator Laboratory, 2575 Sand Hill Road, 
Menlo Park, CA 94205, USA}
\affiliation{Kavli Institute for Particle Astrophysics and Cosmology, 
Stanford University, 452 Lomita Mall, Stanford, CA 94309, USA}

\author{H.M.~Ara\'{u}jo} 
\affiliation{Imperial College London, High Energy Physics, Blackett 
Laboratory, London SW7 2BZ, United Kingdom}

\author{X.~Bai} 
\affiliation{South Dakota School of Mines and Technology, 501 East St 
Joseph St., Rapid City, SD 57701, USA}

\author{A.J.~Bailey} 
\thanks{Corresponding author: a.bailey12@imperial.ac.uk}
\affiliation{Imperial College London, High Energy Physics, Blackett 
Laboratory, London SW7 2BZ, United Kingdom}

\author{J.~Balajthy} 
\affiliation{University of Maryland, Department of Physics, College 
Park, MD 20742, USA}

\author{P.~Beltrame} 
\affiliation{SUPA, School of Physics and Astronomy, University of 
Edinburgh, Edinburgh EH9 3FD, United Kingdom}

\author{E.P.~Bernard} 
\affiliation{Yale University, Department of Physics, 217 Prospect St., 
New Haven, CT 06511, USA}

\author{A.~Bernstein} 
\affiliation{Lawrence Livermore National Laboratory, 7000 East Ave., 
Livermore, CA 94551, USA}

\author{T.P.~Biesiadzinski} 
\affiliation{Case Western Reserve University, Department of Physics, 
10900 Euclid Ave, Cleveland, OH 44106, USA}
\affiliation{SLAC National Accelerator Laboratory, 2575 Sand Hill Road, 
Menlo Park, CA 94205, USA}
\affiliation{Kavli Institute for Particle Astrophysics and Cosmology, 
Stanford University, 452 Lomita Mall, Stanford, CA 94309, USA}

\author{E.M.~Boulton} 
\affiliation{Yale University, Department of Physics, 217 Prospect St., 
New Haven, CT 06511, USA}

% ...

\collaboration{LUX Collaboration}

\date{\today}

\begin{abstract}
We present experimental constraints on the spin-dependent WIMP-nucleon elastic cross sections from LUX data acquired in 2013. LUX is a dual-phase xenon time projection chamber operating at the Sanford Underground Research Facility (Lead, South Dakota), which is designed to observe the recoil signature of galactic WIMPs scattering from xenon nuclei. A profile likelihood ratio analysis of 1.4$\times 10^{4}$~kg$\cdot$days of fiducial exposure allows 90\% CL upper limits to be set on the WIMP-neutron (WIMP-proton) cross section of $\sigma_n$ = 9.4$\times 10^{-41}$~cm$^2$ ($\sigma_p$~=~2.9$\times~10^{-39}$~cm$^2$) at 33~GeV/c$^2$. The spin-dependent WIMP-neutron limit is the most sensitive constraint to date.

\end{abstract}

\pacs{PACS Numbers}

\maketitle


The Weakly Interacting Massive Particle (WIMP) is one of the leading candidates for explaining the observed abundance of dark matter in the Universe \cite{Feng2010}. Astronomical evidence for the existence of dark matter ranges from galactic to cosmological scales \cite{Read2014, Harvey2015, Planck2015}. However, its exact composition remains unknown. WIMPs arise in many extensions of the standard model of particle physics and are expected to have a small coupling to ordinary matter \cite{Drukier1986}. The Large Underground Xenon (LUX) experiment is designed to detect the low-energy scattering of galactic WIMPs with atomic nuclei.

LUX is a dual-phase xenon time projection chamber (TPC) with 250~kg active mass, currently operating at the Sanford Underground Research Facility (SURF) in Lead, South Dakota \cite{Akerib2013}. A WIMP interaction in the detector gives a low energy nuclear recoil ($\lesssim$ 100~keV), producing prompt scintillation light (S1) and ionization electrons. An applied electric field (180~V/cm between the cathode and gate electrodes) drifts the electrons upwards into the gaseous phase of the detector, where they produce electroluminescence (S2). Photons are detected by two arrays of photomultiplier tubes (PMTs). The difference in arrival time between the S1 and S2 signals gives the depth of the interaction, and the $(x,y)$ position is found from the localization of the S2 in the top PMT array. The ability to reconstruct positions of interactions in three dimensions allows fiducialization of the active volume, avoiding higher background regions near the detector walls and enabling rejection of multiple scatters. Electronic recoils (ER) are distinguished from nuclear recoil (NR) interactions by the ratio of the charge (S2) and scintillation (S1) signals.

LUX published world-leading limits on the spin-independent (SI) WIMP-nucleon scattering cross section from an exposure of 1.1$\times 10^{4}$~kg$\cdot$days in 2013 \cite{Akerib2014}, for WIMP masses above 5.7~GeV. After collecting these data, a low energy NR calibration \cite{LUXDD} was performed with a Deuterium-Deuterium (DD) neutron generator. This allows the charge and light response to be evaluated down to 1.1~keV, below the 3~keV recoil energy cut-off imposed in the original analysis. In addition, high statistics ER calibration data were acquired with a tritium source dissolved in the active liquid xenon \cite{LUXTritium}, improving the characterization of the detector response to low energy ER interactions. A further 10~live days of WIMP search data were added taking the exposure up to 95 live days (1.4$\times 10^{4}$~kg$\cdot$days). Other improvements were made to the background model, vertex reconstruction, and event selection. These improvements motivated a reanalysis of the 2013 data, enhancing the sensitivity of the LUX experiment \cite{LUXReanalysis}. The SI result is compatible with the background-only hypothesis and sets a 90\% confidence upper limit on the WIMP-nucleon cross section of 5.6$\times 10^{-46}$~cm$^2$ at a WIMP mass of 33~GeV/c$^2$. Now we use the reanalyzed data to also set limits on the spin-dependent (SD) WIMP-proton and WIMP-neutron scattering cross sections.

The case for an axial-vector (spin-dependent) interaction is well motivated and occurs in various theories beyond the standard model of particle physics, including supersymmetry \cite{Jungman1996}, universal extra dimensions \cite{Hooper2007}, and little Higgs theories \cite{Birkedal-Hansen2004, Hubisz2005}. For the nonrelativistic velocities of galactic WIMPs, the scattering is mostly coherent across the whole nucleus. For isospin-conserving interactions, this leads to an enhancement of the scalar (spin-independent) interaction proportional to $A^2$. For the SD case there is cancellation between the spins of nucleon pairs so the $A^2$ enhancement is not present. Therefore, nuclei with even numbers of protons and neutrons have almost zero nuclear spin, giving a negligible contribution to the SD interaction. However, models exist where the SI interaction is suppressed \cite{Bednyakov1994}, making it essential to search for the SD interaction, and moreover, xenon is sensitive to this interaction because it contains isotopes with nonzero spin.

SI and SD are not the only possible interactions. In a general, nonrelativistic effective field theory treatment there are several possible operators \cite{Fitzpatrick2013}. In particular, there are two ways in which WIMPs can couple to spin, with the projection of the spin either parallel or perpendicular to the momentum transfer. The standard SD response is a linear combination of these, only including operators which are nonvanishing at zero momentum transfer. Constraints on the complete set of operators will be presented in a future publication. 


For direct detection experiments the principal measured quantity is the standard WIMP-nucleus cross section $\sigma_0$. The WIMP-nucleus differential cross section for momentum transfer $q$ for the SD interaction can be written in terms of $\sigma_0$ \cite{Klos2013}:

\begin{equation}
\frac{d\sigma}{dq^2} = \frac{8G_F^2}{(2J+1)v^2} S_A(q) = \frac{\sigma_0}{4\mu_N^2 v^2} \frac{S_A(q)}{S_A(0)} ,
\end{equation}

\noindent where $G_F$ is the Fermi constant, $\mu_N$ is the WIMP-nucleus reduced mass, $J$ is the total nucleus spin, $v$ is the WIMP velocity relative to the target and $S_A$ is the spin structure function. $S_A$ is analogous to the form factor in the SI case; it describes the spin distribution within the nucleus. All momentum dependence is contained in the $S_A(q)$ term. In order to compare direct detection experiments with different target nuclei the WIMP-nucleon cross section is required. For $q=0$, $S_A$ reduces to:

\begin{equation}
\label{SA0}
S_A(0) = \frac{(2J+1)(J+1)}{4\pi J} | (a_0 + a'_1)\langle S_p \rangle + (a_0 - a'_1) \langle S_n \rangle|^2 ,
\end{equation}

\noindent where $\langle S_{p,n} \rangle$ are the proton or neutron spin expectation values averaged over the nucleus and $a_{0,1}$ are the isoscalar and isovector couplings.  These are related to the WIMP couplings to protons and neutrons by $a_0 = a_p + a_n$ and $a_1 = a_p - a_n$. Then, $a'_1 = a_1(1+\delta a_1(0))$ includes the effects of two-body currents in the $\delta$ term,\footnote{Most previous analyses have not included 2-body currents, which simplifies this equation.} which represent couplings between a WIMP and two nucleons \cite{Epelbaum2009}. In this zero-momentum transfer limit we can separate the two cases of ``proton-only'' ($a_0$~=~$a_1$~=~1) or ``neutron-only'' ($a_0$~=~$-a_1$~=~1) couplings and write:

\begin{equation}
\sigma_{p,n} = \frac{3 \mu_{p,n}^2 (2J + 1)}{4 \pi \mu_N^2} \frac{\sigma_0}{S_A(0)} .
\end{equation}


$S_A(q)$ can be obtained from detailed nuclear shell model calculations. The result depends on which nuclear states are included and the allowed configurations of nucleons within those states. There are also differences in the nuclear interactions accounted for. The calculation used here is from Klos {\it et al.} \cite{Klos2013}. It includes the largest number of states and allowed configurations compared to previous theoretical treatments in the literature. The order of the experimentally measured nuclear energy levels in xenon is reproduced well. In addition, the Klos {\it et al.} result uses a chiral effective field theory treatment of the nuclear interactions including two-body currents. These structure functions are an update of those in Ref. \cite{Menendez2012}. Within the recoil energy range of interest, changes to the neutron-only structure function are small: at most 5\% for $^{129}$Xe, and a maximum 20\% increase for $^{131}$Xe. For proton-only the structure function is smaller than previously: as the recoil energy increases the difference in $^{129}$Xe rises to 30\% and in $^{131}$Xe to 50\%. We also compare to the structure function calculation of Ressell and Dean with the Bonn A nucleon-nucleon potential \cite{Ressell1997}, which has been extensively used in previous SD results. This includes the same states as Ref.~\cite{Klos2013}, but has more truncations in the allowed configurations of nucleons and only includes interactions with one nucleon.

There are two naturally occurring xenon isotopes with an odd number of neutrons, $^{129}$Xe and $^{131}$Xe (abundances 29.5\% and 23.7\%, respectively). Therefore, the ``neutron-only'' sensitivity is much higher than ``proton-only'', as the majority of the nuclear spin is carried by the unpaired neutron. When only WIMP interactions with one nucleon are considered, the choice of $a_{p,n}$ above corresponds to WIMPs either coupling to only protons or neutrons. However, once two-body currents are included an interaction between a WIMP, a proton, and the unpaired neutron can occur even in the ``proton-only'' case. Therefore, this gives a significant enhancement to the structure function for ``proton-only'' coupling, while only slightly reducing the ``neutron-only''.

Single scatter events (one S1 followed by one S2) within the fiducial volume (radius $<$ 20~cm, 38--205~$\mu$s drift~time, or 48.6--8.5~cm above bottom PMT faces in z) are selected for the analysis. A total of 591 events are observed in the region of interest (cf. Fig. 2 in Ref.~\cite{LUXReanalysis}) during an exposure of 1.4$\times 10^{4}$~kg$\cdot$days. The background rate originating from NR events is negligible \cite{LUXReanalysis} but ER events produce a significant background. The ER backgrounds include external gamma-rays from detector materials, $^{127}$Xe x-rays, and contaminants in the xenon ($^{85}$Kr, Rn) \cite{Akerib2015}. The tritium dataset allows Monte Carlo simulations \cite{LUXSim2012} to be tuned to ER calibration data, which is then used to generate PDFs (in S1 vs. S2) for these ER backgrounds. Another important background comes from radon daughter decays on the PTFE walls of the TPC, with the tail of the distribution in reconstructed radius extending into the fiducial volume \cite{Lee2015}. In these ``wall events'' some electrons are lost, resulting in a reduced S2 signal, so that many events lie below the signal band in S2/S1. Part of this background is ERs, which can mimic NRs due to their reduced S2 signal. There are also NR wall events from the alpha decay of $^{210}$Po, which produces a recoiling daughter $^{206}$Pb nucleus. The PDF model for the wall events is generated from sidebands in the data.



For SD scattering the signal spectrum (per unit cross section) is suppressed relative to the SI case. The shape of the recoil spectrum produced by a SD neutron-only interaction is very similar to that from a SI one. The SD proton interaction produces a somewhat harder recoil spectrum at all WIMP masses, with the effect growing for heavier WIMPs; at 20 TeV the SD proton-only has 28\% of recoils between 25 and 50 keV, compared to 20\% for SI. The signal PDF for a given WIMP mass is evaluated by fitting the yield of single scatters from the DD-neutron calibration in S2 and S1 \cite{LUXDD}. Systematic uncertainties from the DD neutron calibration are included in this fit. Contributions from the different isotopes are accounted for by adding their differential event rates. Confidence intervals are set with a profile likelihood ratio (PLR) in four variables: S1, S2, radius, and height. All of these variables are useful for discriminating signal from background. Further detail on the analysis can be found in Refs.~\cite{LUXReanalysis, LUXPRD}, including the application of a power constraint at the median sensitivity so as not to benefit from background fluctuations. The observed events are consistent with the expectation from background only.


\begin{figure}
\includegraphics[width=8.5cm]{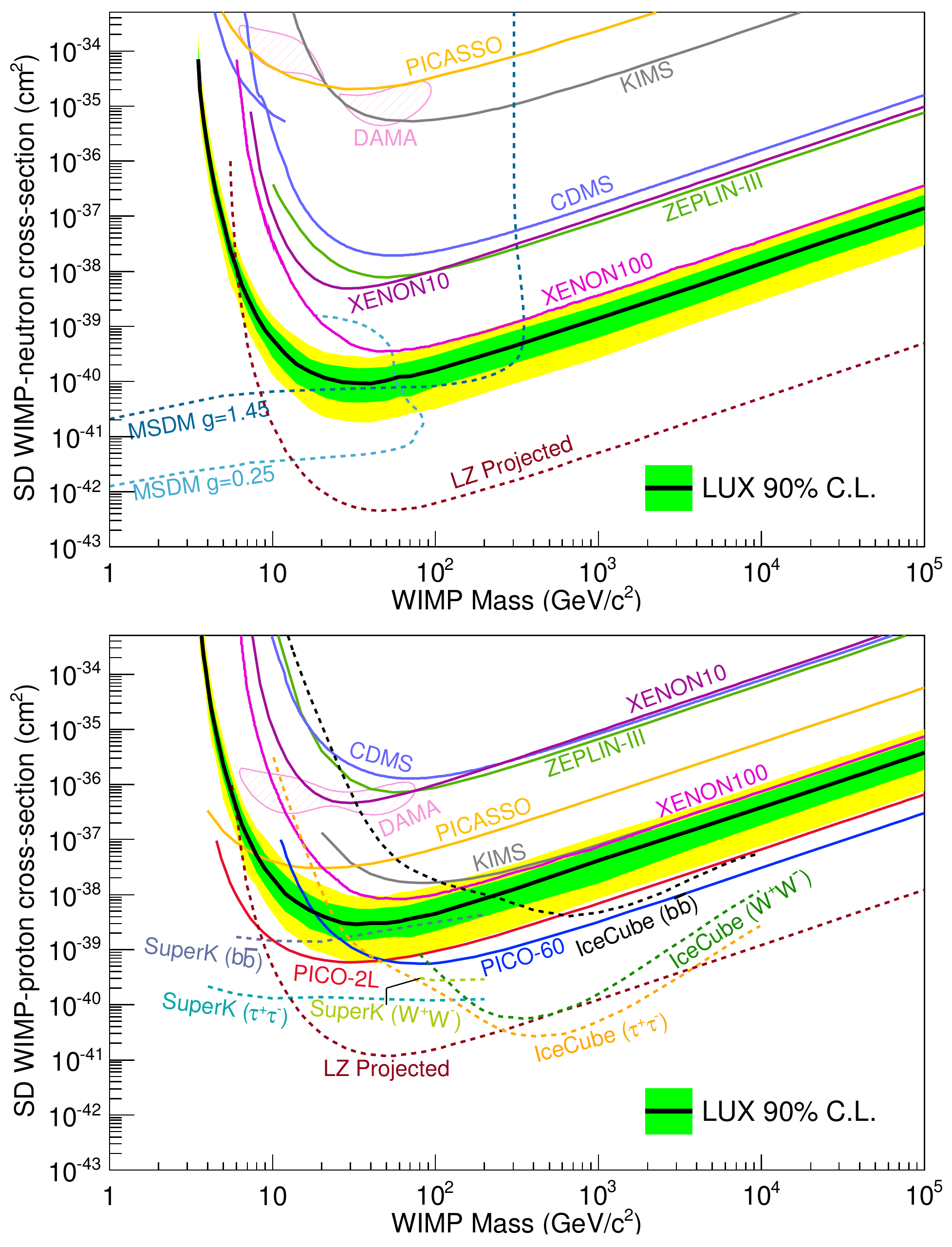}
\caption{\label{Limits} LUX upper limits on the WIMP-neutron (top) and -proton (bottom) elastic SD cross sections at 90\% C.L. The observed limit is shown in black with the $\pm$1$\sigma$ ($\pm$2$\sigma$) band from simulated background-only trials in green (yellow). Also shown are the 90\% C.L. from: CDMS \cite{Ahmed2011}, KIMS \cite{Lee2007, Kim2012}, PICASSO \cite{Archambault2012}, PICO-2L \cite{Amole2016}, PICO-60 \cite{Amole2015_60}, XENON10 \cite{Angle2008}, XENON100 \cite{Aprile2013}, and ZEPLIN-III \cite{Akimov2012,Lebedenko2009}. The DAMA allowed region at 3$\sigma$ as interpreted in \cite{Savage2009} without ion channeling is the shaded areas. Three indirect limits from IceCube \cite{Aartsen2016} and SuperK \cite{Choi2015} are shown. Collider limits from CMS mono-jet searches are included, assuming the MSDM model with two coupling scenarios \cite{Malik2015}. The projected sensitivity for the LZ experiment is shown for an exposure of 5.6$\times 10^{5}$~kg$\cdot$days \cite{LZ_CDR}.}
\end{figure}

The upper limits on the SD WIMP-nucleon cross sections from the PLR analysis are shown in Fig.~\ref{Limits}. The minimum excluded cross section at 90\% CL for WIMP-neutron (WIMP-proton) elastic scattering is $\sigma_n$~=~9.4$\times 10^{-41}$~cm$^2$ ($\sigma_p$~=~2.9$\times 10^{-39}$~cm$^2$), for a WIMP mass of 33~GeV/c$^2$. For the neutron-only coupling the excluded cross section is lower than from previous direct searches. The proton-only limit is less constraining by a factor of {\raise.17ex\hbox{$\scriptstyle\sim$}}$30$. Using alternative structure functions from Ref.~\cite{Ressell1997}, the neutron-only upper limit is improved by a factor \raise.17ex\hbox{$\scriptstyle\sim$}$0.5$ and the proton-only degraded by \raise.17ex\hbox{$\scriptstyle\sim$}$2.5$. The results presented here improve on the limits set in Ref.~\cite{Savage2015} owing mostly to the lower energy threshold and the better background rejection afforded by the PLR-based statistical analysis. PICO \cite{Amole2016, Amole2015_60} is more sensitive to proton-only coupling, due to the unpaired proton of the fluorine nuclei in the C$_3$F$_8$ target. However, the inclusion of two-body currents in the xenon structure functions yields significant proton-only sensitivity and the proton-only limit from this result is competitive. The DAMA allowed region \cite{Savage2009} is excluded even in the proton-only case by this result.

Collider searches for dark matter particles can be interpreted in the same parameter space as direct searches for particular conditions  \cite{Malik2015}. In Fig.~\ref{Limits} we include limits from CMS mono-jet searches \cite{CMS2015}, assuming the Minimal Simplified Dark Matter (MSDM) model for the particular case where the couplings of the mediator to the quarks and the dark matter particle are equal ($g=g_q=g_{DM}$). The cross section is dependent on these couplings, so we compare to the smallest and largest values used in Ref.~\cite{Malik2015}. For low WIMP masses the collider limits are stronger for both couplings, but these searches are not sensitive to heavier WIMPs. It is important to note this interpretation of collider searches is model-dependent. Therefore, dark matter signals would ideally be observed in collider, indirect, and direct searches in order to fully investigate the interactions of WIMPs.

\begin{figure*}
\includegraphics[width=11.0cm]{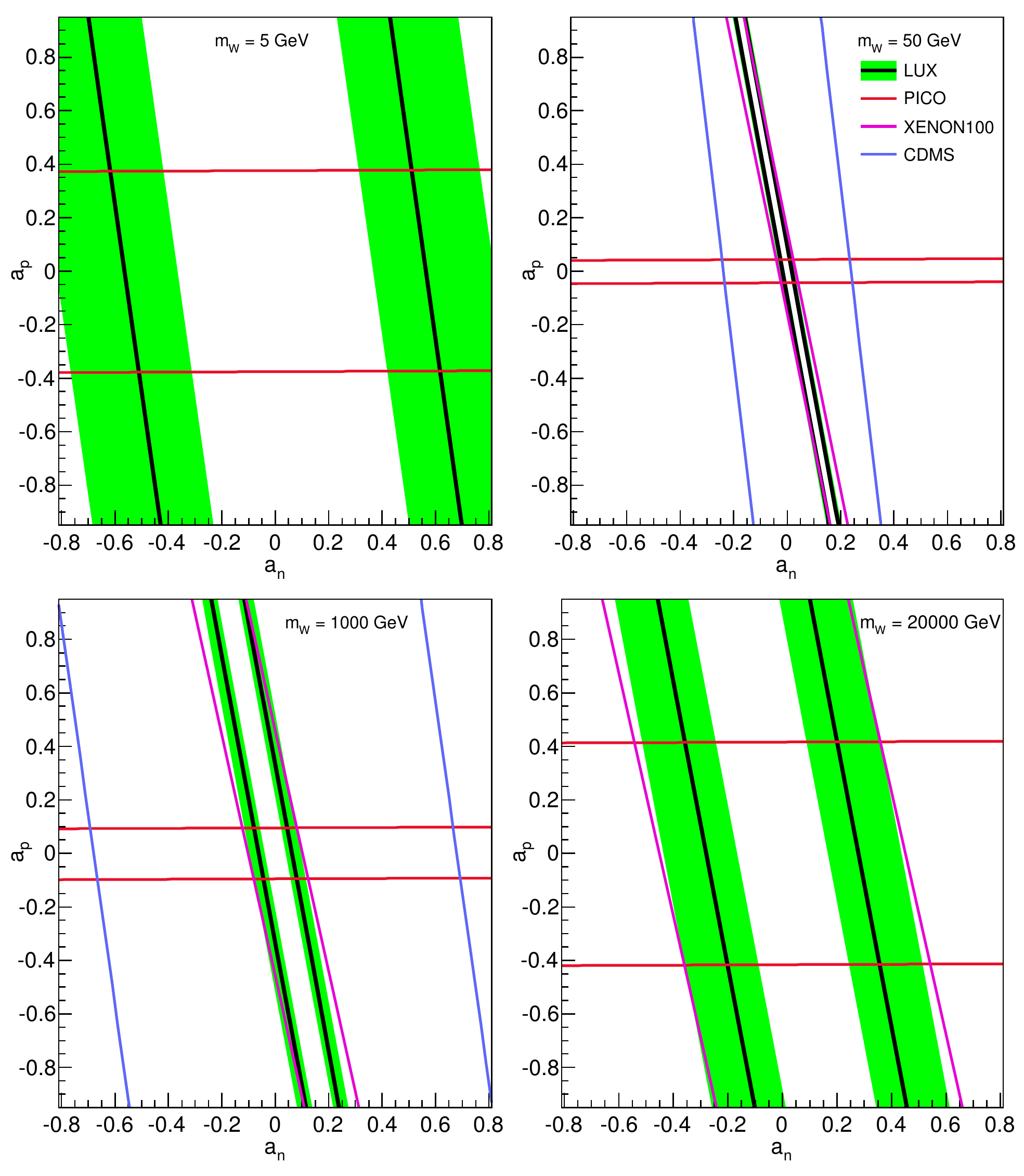}
\caption{\label{ap_an_plot} Constraints on the effective WIMP couplings to protons and neutrons, $a_p$ and $a_n$, at 90\% CL for various WIMP masses (5, 50, 1000, 20000 GeV). Also shown are CDMS \cite{Ahmed2009}, PICO-2L (for 5~GeV) \cite{Amole2016}, PICO-60 (for other masses) \cite{Amole2015_60} and XENON100 \cite{Aprile2013}, where the constraints have been inferred from the limits on $\sigma_{p,n}$ using the method in Ref.~\cite{Giuliani2005}.}
\end{figure*}

With limits set on $\sigma_{p,n}$ the allowed region in $a_p-a_n$ space can be found following the procedure detailed in \cite{Tovey2000}:

\begin{equation}
\sum_A \left( \frac{a_p}{\sqrt{\sigma_p^{A}}} \pm \frac{a_n}{\sqrt{\sigma_n^{A}}} \right)^2 > \frac{\pi}{24 G_F^2 \mu_p^2} ,
\end{equation}

\noindent where $\sigma_{p,n}^A$ are the limits on the proton or neutron-only cross sections, for the isotope with mass number $A$. The excluded region is shown in Fig.~\ref{ap_an_plot}. Typically only the most sensitive channel of the two cross sections is shown. In this case the limits in the $a_p-a_n$ plane can be found following the method detailed in Ref.~\cite{Giuliani2005}, which is a good approximation if $a_p \gg a_n$ or vice-versa.

This result improves the constraint on $a_n$ over previous experiments. The lines are parts of elongated ellipses and the orientation depends on the sensitivity to both $a_p$ and $a_n$. The angle of the ellipse for LUX and XENON100 is not the same due to differences in the spin structure functions used and the energy scale in the analysis (which affects the signal spectrum). XENON100 also had slightly different abundances of $^{129}$Xe and $^{131}$Xe, due to the addition of isotopically modified xenon. This plot also emphasizes the complementarity between the different detector materials. 

In conclusion, we have set the most stringent limits on the SD WIMP-neutron cross section for all WIMP masses down to 3.5~GeV/c$^2$ from the 2013 LUX data, and the proton-only limit is also competitive. We also improve the constraints on the possible values of the couplings $a_p$ and $a_n$, complementary to experiments that are more sensitive to the proton than the neutron coupling. The sensitivity to both proton and neutron-only coupling will be improved greatly with future large-scale experiments with xenon targets such as LZ \cite{LZ_CDR}.

\begin{acknowledgments}
This work was partially supported by the U.S. Department of Energy (DOE) under award numbers DE-FG02-08ER41549, DE-FG02-91ER40688, DE-FG02-95ER40917, DE-FG02-91ER40674, DE-NA0000979, DE-FG02-11ER41738, DE-SC0006605, DE-AC02-05CH11231, DE-AC52-07NA27344, and DE-FG01-91ER40618; the U.S. National Science Foundation under award numbers PHYS-0750671, PHY-0801536, PHY-1004661, PHY-1102470, PHY-1003660, PHY-1312561, PHY-1347449, PHY-1505868; the Research Corporation grant RA0350; the Center for Ultra-low Background Experiments in the Dakotas (CUBED); and the South Dakota School of Mines and Technology (SDSMT). LIP-Coimbra acknowledges funding from Funda\c{c}\~{a}o para a Ci\^{e}ncia e a Tecnologia (FCT) through the project grant PTDC/FIS-NUC/1525/2014. Imperial College and Brown University thank the UK Royal Society for travel funds under the International Exchange Scheme (IE120804). The UK groups acknowledge institutional support from Imperial College London, University College London and Edinburgh University, and from the Science \& Technology Facilities Council for PhD studentships ST/K502042/1 (AB), ST/K502406/1 (SS) and ST/M503538/1 (KY). The University of Edinburgh is a charitable body, registered in Scotland, with registration number SC005336.

This research was conducted using computational resources and services at the Center for Computation and Visualization, Brown University.

We gratefully acknowledge the logistical and technical support and the access to laboratory infrastructure provided to us by the Sanford Underground Research Facility (SURF) and its personnel at Lead, South Dakota. SURF was developed by the South Dakota Science and Technology Authority, with an important philanthropic donation from T. Denny Sanford, and is operated by the Lawrence Berkeley National Laboratory for the Department of Energy, Office of High Energy Physics.
\end{acknowledgments}
%

\bibliography{MyCollection}

\end{document}